# Large-eddy simulations of geophysical turbulent flows with applications to planetary boundary layer research

*Igor Esau[1]*

[1] Nansen Environmental and Remote Sensing Center, Thormohlensgate 47, 5006, Bergen; e-mail igore@nersc.no

**Abstract**

The present study gives an overview and emphasizes principal moments of the applications of the turbulence-resolving modeling with large-eddy simulation (LES) numerical technique to planetary boundary layer (PBL) research and climate studies. LES have been applied to the PBL research since its emergence at the end of 1960s. LES proved to be very useful in understanding of the atmospheric and ocean turbulent exchange and ultimately in parameterization improvement in traditional meteorological models. LES have played a key role in recognizing the importance of previously ignored self-organized structures in the geophysical turbulence. LES assisted theoreticians and weather/climate modelers with reliable information about the averaged vertical structure of the PBL in convection and shear regimes as well as with better estimations of key PBL parameters, e.g. an entrainment rate, for model calibrations.

At present, LES are an essential, indispensible part of geosciences, while the mainstream of the LES research still deals with idealized case studies with rather simple micro-physics. Advance in computer performance and parallel computations opens a way to more realistic studies such as coupled studies of the atmosphere and ocean turbulence and the urban PBL. Moreover, simplified LES sub-module is run as a kind of "super-parameterization" in a global circulation model. It radically improves representation of the convective clouds in the model, leading to much better understanding and prediction of hurricanes and possibly climate features.

Through better understanding of the high Reynolds number PBL flows, the LES improve themselves. LES experience and following observational campaigns e.g. the HATS experiment disclosed necessary and sufficient conditions to conduct reliable and adequate simulations. It has been demonstrated that contrary to earlier believes, neither the unresolved surface sub-layer nor strong static stability are preventing robust LES experiments.

# 1. INTRODUCTION

Geophysical flows are often turbulent. This is especially true for the flows near the land-atmosphere/ocean interface where large gradients of meteorological quantities such as wind speed and direction and temperature are observed. This lower turbulent layer of the atmosphere is called the planetary boundary layer (PBL). Similar layer under the ocean surface is called the ocean mixed layer (OML) and the turbulent layer above the ocean bottom is called the bottom boundary layer. Unless it is specially articulated we will consider only the PBL. Interested reader may refer to e.g. Garratt (1992) and Baumert et al. (2005) for more systematic physical introduction in the PBL and the OML respectively.

Formidable difficulties of the PBL research are related to spatial and time scales of the PBL as a physical object. A typical PBL is of $h = O(10^2$ m$)$ deep, where $h$ is the PBL depth measured as the thickness of the essentially turbulent layer (e.g. Zilitinkevich et al., 2007). Moreover, the satellite and aerial images reveal that the PBL turbulence is organized, at least on the largest scales, in persistent pattern of quasi-regular spatial structures with a typical horizontal scale of $\lambda \sim 2-10h$ (e.g. Atkinson and Zhang, 1996). In the case of surface thermal or roughness heterogeneity, internal PBL, larger scale gravity flows and gravity waves develop as well. Their interactions add significant complicity into the PBL dynamics (e.g. Esau, 2007). The scale of the PBL leads to very large Reynolds number, Re, of the flow, which is typically of Re $= O(10^6)$. At the same time Richardson and Rossby numbers are within a sensitive range for the PBL turbulence.

The large scales of the PBL turbulence result in important difference between geophysical and engineering fluid mechanics. The PBL cannot be observed directly throughout the entire turbulent layer leave alone in several locations. Typical height of a meteorological mast is of $O(10$ m$)$ with just a few masts (e.g. in Colorado, USA; Cabaw, the Netherlands; and Obninsk, Russia), which are of $O(300$ m$) \sim O(h)$ height. The large masts have own drawbacks. They perturb flow considerably. In principle, remote sensing methods, e.g. LIDAR, SODAR and similar instruments, could help observations but at present neither their range nor accuracy are satisfactory.

The best PBL data so far can be collected during expensive and labor intensive turbulence campaigns where high-quality frequent radio-soundings, air-born and ground-born measurements are combined. An example of such a campaign is the SHEBA (Surface Heat and Energy Balance of the Arctic Ocean) experiment conducted in 1997-1998 (Uttal et al., 2002). There is another problem however. Such observations are limited in time and hence their scientific outcome depends very much on the luck with weather and data acquisition.

To end with troubles of PBL researchers, there is a problem with controllability and repeatability of external conditions governing the PBL evolution. Often, parameters critical for the PBL understanding have not been measured properly during the campaign. The following measurements cannot be repeated under similar conditions. It leads to large, irreducible uncertainty in the data that allow for alternative theories. These difficulties are exemplified by the studies of the imposed atmospheric stability effect (Zilitinkevich and Esau, 2005; Esau and Zilitinkevich, 2006). The stability effect on the

PBL has been suspected for long time (Byun, 1991; Overland and Davidson, 1992) but only LES experiments identified the parameters to look at. In fact, LES was a crucial tool in development of a constructive theory of the PBL sensitivity to the imposed stability.

LES provides certain advantages. LES are accurate, reliable, controllable source of data. The LES experiments can be easily repeated under exactly the same or modified conditions. The LES data are regular in space and time comprising the entire PBL and, at least in principle, adequate to spatial and time scales of the PBL dynamics. All these make LES very attractive for the PBL research. It is worth noticing that there are rather deep philosophical as well as technical problems related to LES applications in geophysics (e.g. Stevens and Lenshow, 2001).

This paper is devoted to overview of the LES advantages, problems and intrinsic limitations having in focus the geophysical, meteorological at the first place, applications of this numerical technique. It is worth mentioning advantages of LES with respect to other competing numerical techniques, namely, direct numerical simulations (DNS) and Reynolds averaged numerical simulations (RANS). LES is a kind of the best compromise between computationally heavy DNS and over-assumed RANS. As known (e.g. Lesieur, 1997) the number of operations required for DNS scales as $2^m \operatorname{Re}^{\frac{3}{4}m}$ where $m = 3$ is the physical space dimension. The number of operations in LES scales less dramatic as $2^m \operatorname{Re}^{\frac{1}{2}m}$. Thus, at Re = $10^6$, LES is about $10^{9/2}$ or by factor of 100,000 more efficient and requires about 10,000 times less memory than DNS. This efficiency is achieved through neglecting details of turbulent energy cascade in the large part of inertial and the whole dissipation sub-ranges of scales. Fortunately, observations reveal that the inertial sub-range of scales spans several decades in wave numbers in the PBL so that simulations with the mesh resolution of $O(10\ m)$ are well within the universal inertial sub-range of scales. Another problem with DNS is difficulty to reproduce the surface roughness, which is in practical cases expressed through an approximate roughness length in the logarithmic law of the wall. The most popular approach and the least computationally expensive is RANS. In principle, well resolved LES could be run with RANS used as sub-grid turbulence closure (e.g. Schumann, 1989; Schumann and Moeng, 1991; Mestayer, 1996). This is due to the very simple reason that the role of the closure in LES is relatively small (Meyers et al., 2003; Meyers and Baelmans, 2004). In fact, LES could be considered as DNS of a special non-linear fluid, so called LES-fluid, far from boundaries and under weak stratification of the flow (Mason, 1994; Muschinski, 1996; Germano, 2002). However, geophysical LES is often under-resolved in order to compromise the large scale difference between the self-organized and inertial-scale turbulence. External parameters such as surface heterogeneity scale also affect the LES scales.

The structure of the paper is as follows. The next Section 2 discusses correspondence between LES data and observed PBL turbulence. Section 3 considers the simplest, neutrally stratified PBL or the Ekman boundary layer (EBL). Here, the surface layer problems and emergence of the self-organized turbulence are brought to the reader's attention. Section 4 is devoted to more applied geophysical LES studies. The last Section 5 summarizes unsolved problems and prospective LES.

## 2. CORRESPONDENCE BETWEEN LES AND GEOPHYSICAL TURBULENCE

Large-eddy simulation (LES) is a numerical technique aimed on three-dimensional direct numerical simulations of relatively large-scale turbulent fluctuations in a flow whereas small-scale fluctuations are parameterized. The meaning of the term "large-scale" in geophysical applications has to be correctly understood. Figure 1 shows the observed non-dimensional energy spectra in the upper surface layer of the PBL.

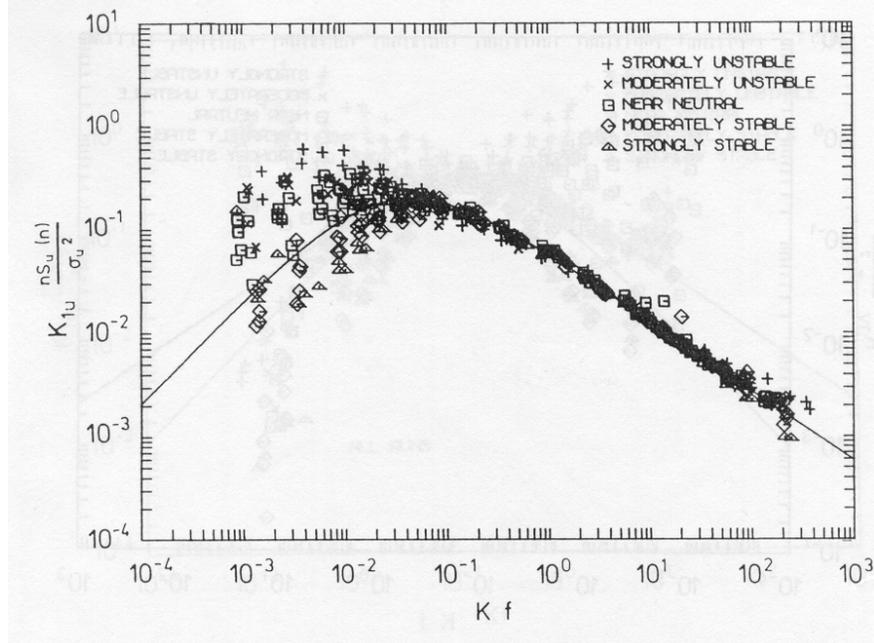

**Fig.1:** Observed non-dimensional energy spectra in the upper surface layer over Baltic Sea. Reproduced with permission from Larsen et al. (1985).

As it is obvious, the PBL turbulence spectra have two distinct sub-ranges, namely, the universal scaling inertial sub-range where the turbulent spectra collapse well and low-frequency, non-universal sub-range where such a collapse is not observed. Statistical properties of the inertial sub-range turbulence are well known[1] and generally out of interest for geophysical applications. This lack of interest to the inertial sub-range is easy to explain from Fig. 1. The plot is given in log-log scale. Hence, the actual root-mean square fluctuation velocity, $u' = (2E)^{1/2} = (\sigma_u^2 + \sigma_v^2 + \sigma_w^2)^{1/2} = O(0.1 \text{ m s}^{-1})$, in the inertial sub-range is just a small fraction of the typically observed $u' = O(1 \text{ m s}^{-1})$. The deeper in the inertial sub-range we move the smaller $u'$ could be found as the turbulent kinetic energy (TKE) scales as $E = C_k \varepsilon^{2/3} k^{-5/3}$ where $\varepsilon$ is the energy dissipation rate and $C_k$ is the Kolmogorov's constant. Then it is not surprising that applications require only representation of the uppermost part of this sub-range.

---

[1] It needs to be mentioned that in stratified flows the classical Kolmogorov inertial sub-range could be modified by the buoyancy force. It results in another Borgiano-Obukhov scaling. Degree to which this modification is important is debated (e.g. Antolelli et al., 2003). We leave these debates out of scopes of this review.

**Correspondence for the turbulence in the inertial sub-range of scales**

In LES, it is important however to represent correctly the spectral energy cascade (spectral flux, $dE/dk$) into the inertial sub-range. This spectral flux ultimately defines the energy dissipation rate. In the quasi steady state high Re PBL, $\varepsilon$ should be on average equal to the energy flux from the largest resolved scales or the mean flow. Thus, in this case, the correct energy flux in the inertial sub-range is the most important internal parameter to be represented in LES. If a part of the inertial sub-range is resolved in LES, the control on $dE/dk$ and $\varepsilon$ can be achieved in a natural manner through a variational formulation of the closure problem. Figure 2 gives the sketch of the problem.

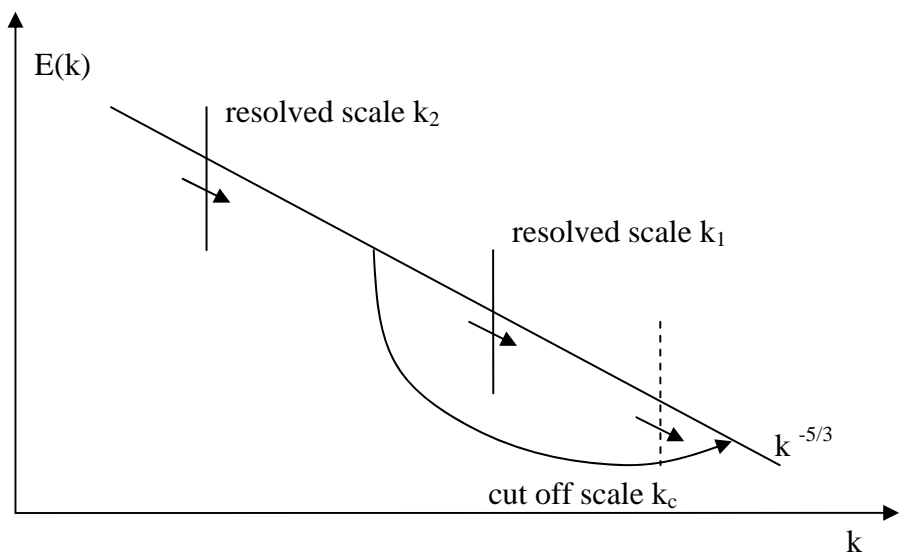

**Fig.2:** A sketch of turbulence closure in LES based on the analysis of the resolved spectral energy flux. The arrows symbolize that $dE/dk$ across any scale in the inertial sub-range are approximately equal.

Since a part of the inertial sub-range is resolved, it is possible to compute $dE/dk$ across the cut off scale, $k_c$, and therefore the energy dissipation rate locally and instantly. Indeed, since the dissipation $\varepsilon = 0$ in the inertial sub-range of scales in the theory of homogeneous isotropic turbulence, the following equality should be satisfied for each fluid volume sufficiently large to contain a spectral shell with $k > k_2$, where $k_2$ is in the inertial sub-range of scales,

$$\frac{\partial E}{\partial k}\bigg|_{k=(k_2+k_1)/2} = \frac{\partial E}{\partial k}\bigg|_{k=k_c} \approx \frac{E(k_2) - E(k_1)}{k_2 - k_1}. \tag{1}$$

Therefore the energy cascade across the shortest resolved scale can be determined through the energy at two sufficiently small resolved scales within the inertial sub-range of scales. For low Re flows, it could be done quite accurately as the spectral interactions are largely local (Bardina et al., 1980). For high Re flows, and moreover, for anisotropic flows, the non-local spectral interactions are essential. The approximation in Eq. (1) is not accurate any longer and a kind of variational problem has to be considered to preserve the global energy conservation and therefore numerical stability of the model. One example of such a problem is to optimize the energy cascade through a certain resolved scale, usually $(2\Delta + \Delta)/2$ where $\Delta$ is the length scale corresponding to the model $k_c$. In this approach, usually called a dynamic Smagorinsky model, all energy at $E((2\Delta + \Delta)/2)$, which excess the amount of energy associated with the inertial energy transfer, should be dissipated instantly by the Smagorinsky eddy-viscosity model with dissipation given as

$$\varepsilon_s = (l_s^2 |S_{ij}|)\left(\frac{\partial u_i}{\partial x_j}\right)^2, \text{ where } S_{ij} = \frac{1}{2}\left(\frac{\partial u_i}{\partial x_j} + \frac{\partial u_j}{\partial x_i}\right). \tag{2}$$

Here the optimization parameter is a mixing length scale $l_s^2$, which is calculated to satisfy Eq. (1), usually written in the physical phase space. Understandably, there are zillion ways to construct the variational problem minimizing energy build up at the smallest resolved scales with general formulation

$$\min\left(\frac{\partial E}{\partial k}|_{k_2 < k_1 < k_c} + \varepsilon_b\right) = 0. \tag{3}$$

Here, both the resolved energy cascade $\frac{\partial E}{\partial k}|_{k_2 < k_1 < k_c}$ and the background dissipation $\varepsilon_b$ take method dependent values.

One robust and simple choice is $\varepsilon_b = \varepsilon_s$ or the Smagorinsky-Lilly eddy viscosity model (Smagorinsky, 1963; Lilly, 1967). If the energy cascade is computed through the Germano equality (Germano et al., 1991) and the mixing length scale $l_s$ varied to satisfy Eq. (3), one ends up with a multitude of dynamic turbulence closures to be used in LES (e.g. Porte-Agel et al., 2000; Zikanov, 2003; Esau, 2004).

The dynamic closures are a kind of standard in engineering LES applications at present. Since the dynamic closures in LES are rather computationally heavy and in some cases hard to implement without the code destabilization, majority of geophysical LES still run the static closure given in Eq. (2) with fixed constant mixing length scale $l_s$. Recently significant efforts have been put to investigate the importance and feasibility of the dynamic closures in PBL simulations. Sullivan et al. (2003) reported analysis of the three-dimensional turbulence data collected during the Horizontal Array Turbulence Study (HATS) field program. The HATS purpose was to reconstruct sub-grid fluxes and variances that are parameterized in LES. Detailed analysis of the motions shows that the ratio between the spectral peak scale, $\Lambda_w$, of the vertical component of velocity fluctuations to the mesh cut off (or filter) scale, $\Delta$, contains the essential information allows for connecting the measured turbulence with its LES approximation. The flux decomposition into the mean modified-Leonard-, $\langle L \rangle$, cross-, $\langle C \rangle$, and Reynolds-, $\langle R \rangle$, terms revealed that these terms are of comparable magnitude at fine ratios, $\Lambda_w / \Delta$ where $\Lambda_w \gg \Delta$. But the flux is dominated by the Reynolds term, or what is the same cannot be reconstructed from the variational problem in Eq. (3) and must be parameterized in a phenomenological way through the approach in Eq. (2), when the turbulence is under-resolved or $\Lambda_w < \Delta$. This is related to the non-local interactions with direct energy transfer to small-scale turbulence but not to the energy backscatter from small-scale turbulence as it has been previously assumed (Mason and Thomson, 1992). The backscatter was found to be less than 20% of the total energy cascade. Figure 3 presents the HATS results.

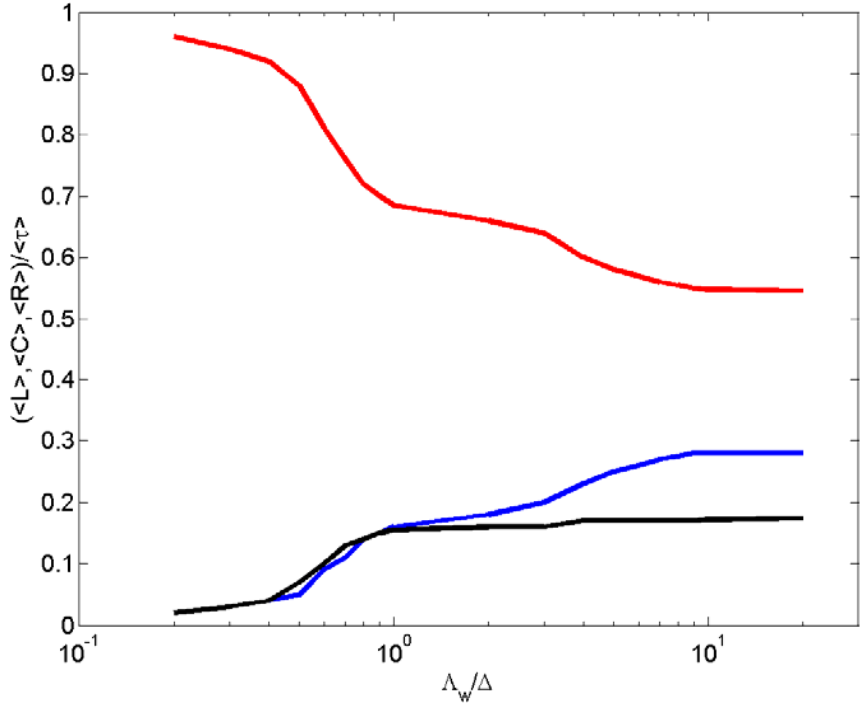

**Fig. 3:** The flux decomposition into the mean modified-Leonard-, $\langle L \rangle$ (blue curve), cross-, $\langle C \rangle$ (black curve), and Reynolds-, $\langle R \rangle$ (red curve), terms obtained in HATS field experiment. The plot is based on data from Figures 9-11 in Sullivan et al. (2003). The flux is presented versus non-dimensional scale ratio between the spectral peak scale, $\Lambda_w$, of the vertical component of velocity fluctuations to the mesh cut off (or filter) scale $\Delta$.

The HATS experiment indicated that the mixing length scale $l_s$ is not a constant in the PBL. It depends on the scale of the TKE spectral maxima. The latter, represented here through $\Lambda_w$, depends on imposed restrictions on the turbulent eddies, namely, on the distance to the surface, $z$, density stratification and the velocity shear profile. The mixing length, $l_s = C_s \Delta$, is often expressed in LES through the Smagorinsky coefficient $C_s$ (Lilly, 1967). Theoretical studies (e.g. Leslie and Quarini, 1979) suggested $C_s \sim 0.18$ to be a universal constant in the homogeneous isotropic non-stratified turbulence. Originally, this values of $C_s$ was adopted in turbulence closures in the geophysical LES (e.g. Deardorff, 1972; Andren, 1994). Numerous experiments and intercomparisons have demonstrated unsatisfactory performance of such LES in geophysical applications, especially those with significant mean wind (e.g. Sullivan et al., 1994).

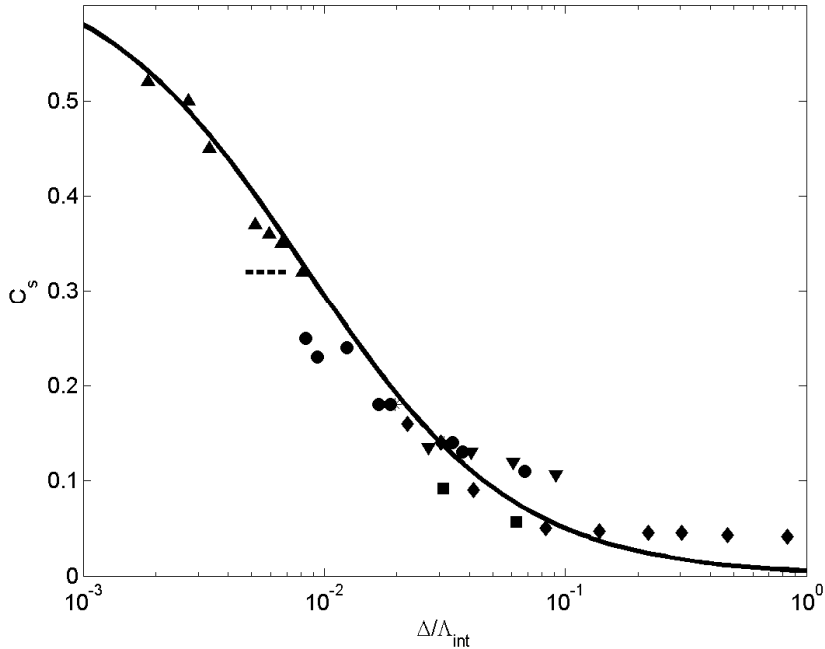

**Fig.4:** Dependence of the Smagorinsky coefficient $C_s$ on the numerical resolution of LES relative to the integral scale of the turbulence $\Lambda_{int}$. The integral scale is taken here to be equal to the scale of the TKE spectral maximum. The symbols are: circles – LESNIC data for the EBL; squares – LESNIC data for homogeneous turbulence; upward triangles – Liu et al. (1994) data; downward triangles – Porte-Agel et al. (2000) data; diamonds – Meneveau (1994) data; asterisk – Emmerich and McGrattan (1998) data; dashed line – typical range in geophysical models (Mason and Brown, 1999); solid line – the best fit approximation after Eq. (4).

It became obvious that $C_s$ is variable. Moreover, generally, it needs be reduced to about $C_s \sim 0.12$ or even less in some cases. The rules controlling $C_s$ were however discovered only very recently after HATS (Horst et al., 2004) and intercomparisons experiments with dynamic LES (Porte-Agel et al., 2000; Esau, 2004; Beare et al., 2006). Figure 4 shows the variability of $C_s$ in LES experiments with LESNIC code (Esau, 2004; Esau and Zilitinkevich, 2006). The best fit curve is

$$C_s = 0.65\left(1 + 120\frac{\Delta}{\Lambda_{int}}\right)^{-1}. \qquad (4)$$

This is in good agreement with HATS findings. It clearly shows that the under-resolved turbulence need to be supported through reduced numerical stability of the model. Similar conclusion follows from Mason and Brown (1999) study of the LES sensitivity to $C_s$ variations. They saw $C_s$ as a measure of numerical accuracy. $\Lambda_{int}$ is a function of stability. It shifts toward smaller scales near the surface and in stably stratified flows. Hence It has been confirmed in HATS (Kleissl et al., 2003; 2004) as well as in dynamic LES (Esau and Zilitinkevich, 2006; Beare et al., 2006). Near the surface, the mixing length scale is following the well-known Prandtl limit (Esau and Byrkjedal, 2007; Zilitinkevich and Esau, 2007) as

$l_s = \kappa z$, where $\kappa$ is the von Karman constant. (5)

Then one can obtain a simple estimation

$$C_s = \frac{l_s}{\Delta} = \kappa \frac{z}{\Delta}. \quad (6)$$

This behavior is in good agreement with what has been observed and simulated (Porte-Agel et al., 2000).

The above overview of small-scale properties of the turbulence and their control role in LES has an important consequence for geophysical applications. It explains early success of the convective PBL LES where $\Delta / \Lambda_{int}$ is small and therefore the assumption $C_s = $ 0.18–0.23 is justified. Similarly it explains enormous difficulties encountered running Ekman layer and even more so stably stratified PBL LES with static turbulence closures. In the former case, the main problem was confined near the surface (Andren et al., 1994) where to simulate correctly the bulk EBL properties the surface layer turbulence has to be over-amplified. The problem become more severe but confined in a thinner layer when the LES resolution increases (Mason, 1994). In the latter case, the turbulence on a coarse mesh had to be maintained with an external artificial energy input in form of random fluctuations (Mason and Thomson, 1992) or waves (Mason and Derbyshire, 1990) or residual large eddies from spin-up period (Saiki et al., 2000). Figure 5 illustrates the difference in LES velocity fluctuations obtained in dynamic-mixed and static Smagorinsky runs of the LESNIC code.

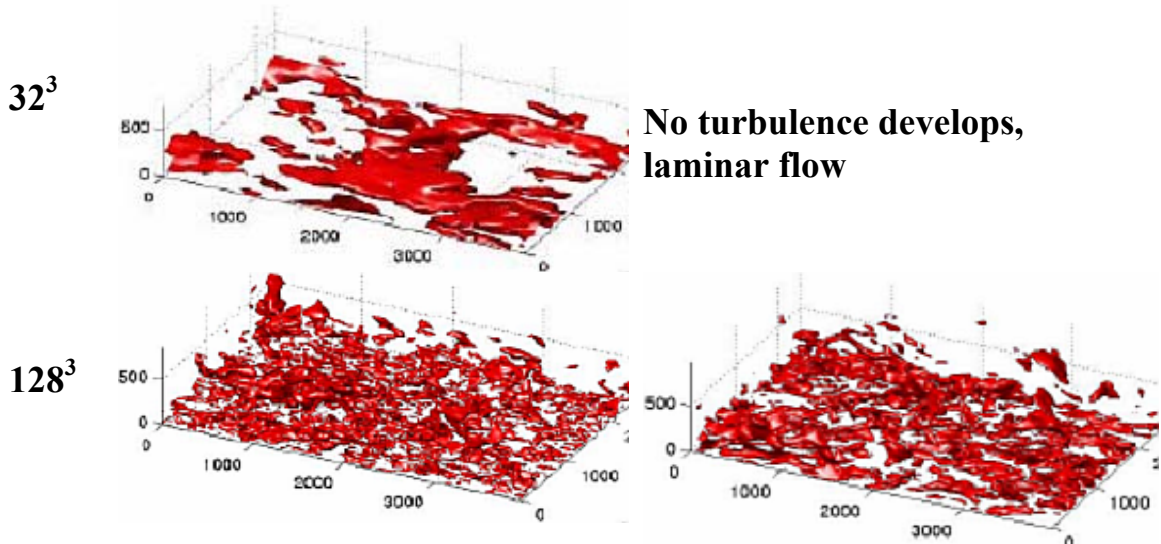

**Fig. 5:** Instant fluctuations of the flow velocity in stably stratified LES runs for Beare et al. (2006) study. Data computed with the LESNIC code (Esau, 2004).

The above overview disclosed the role of correct simulations of the energy cascade in the inertial sub-range of scales for the success of the geophysical LES. At present, it is understood rather clearly that the variational formulation of the turbulence closure, which is not necessary limited to the so-called dynamic eddy-viscosity models, is a key for high quality applied LES. To gain this understanding, the geophysical LES community has adopted results of the computational fluid dynamics and engineering LES communities with certain reservations. At very large geophysical Re, many fine details of this energy cascade are of lesser importance than they were assigned for low Re studies. In particular, although there are already several LES codes with different kind of dynamic closures (e.g. Porte-Agel et al., 2000; Zikanov et al., 2003; Esau, 2004), there are no clear indication that the more sophisticated closure produce more accurate results. To my opinion, there is currently no reason to make the closure more sophisticated than the dynamic Smagorinsky or at the best the dynamic-mixed closure model.

**Correspondence for the turbulence in the energy-containing sub-range of scales**

The correspondence between the geophysical turbulence and their LES counterpart in the energy-containing sub-range of scales is of primary concern for the geophysical community. Despite of this, it has been studied much less than the correspondence on small-scales. This is probably because of somewhat reduced relevance of this problem to the fluid dynamics community. The energy-containing sub-range of scales or the large-scale turbulence (LST) in the PBL is usually limited to 100 m – 1000 m where the upper limit is imposed by the depth of the PBL. The importance of this sub-range for practical applications follows from the energy of fluctuations on these scales and persistence of those fluctuations. The energy and spectral characteristics of the small-scale turbulence is more or less understood from single point observations. The persistence and spatial organization of the LST are less understood. Only recently, Hutchins and Marusic (2007) have reported observations of a very long LST structure in the surface stress qualitatively similar to the streaks in laboratory sheared flows. The geophysical LES are proved to be indispensible to study those characteristics since pioneering works by James Deardorff (Deardorff, 1972).

The LST structure is easy to observe qualitatively on aerial or satellite images when the upper part of those structures is visualized with clouds. For instance, Figure 6 shows an aerial photo of turbulent rolls or "cloud streets" as they are known for meteorologists. Quantitative analysis of the LST in the observations is however difficult. Clouds give imperfect visualization as they are dependent on many non-turbulent factors. Moreover, the random component of the LST makes their identification and automatic quantification rather ambiguous. Finally, the LST characteristics are practically never observed in conjunction with accurate turbulence measurements of fully known meteorological conditions.

Nevertheless, the attempts for compare LST in LES and observations have been published. Mayor et al. (2003) used volume imaging LIDAR observations of an intense and spatially evolving convective PBL on 13 January 1998 during the Lake-Induced Convection Experiment (Lake-ICE). For this comparison, the aerosol scattering was estimated from LES output of relative humidity, a passive tracer, and liquid water. The analysis is based on evaluation of correlation functions of the observed and simulated aerosol backscatter as a function of altitude and offshore distance. The best-fit ellipse approximation of the two-dimensional correlation functions are used to obtain the mean ellipticity and orientation of the structures. The LES and observed correlation functions compared reasonably well throughout the convective PBL including the surface layer where they are expected to disagree. To authors' surprise however, LES performed worse in the middle of PBL where LST are the largest and LES are expected to be robust. As we will see later on, the failure could be related to inadequate accommodation of the LST scales in the limited size domain of LES. The problem of too small domain size is common while probably unavoidable problem of the geophysical LES.

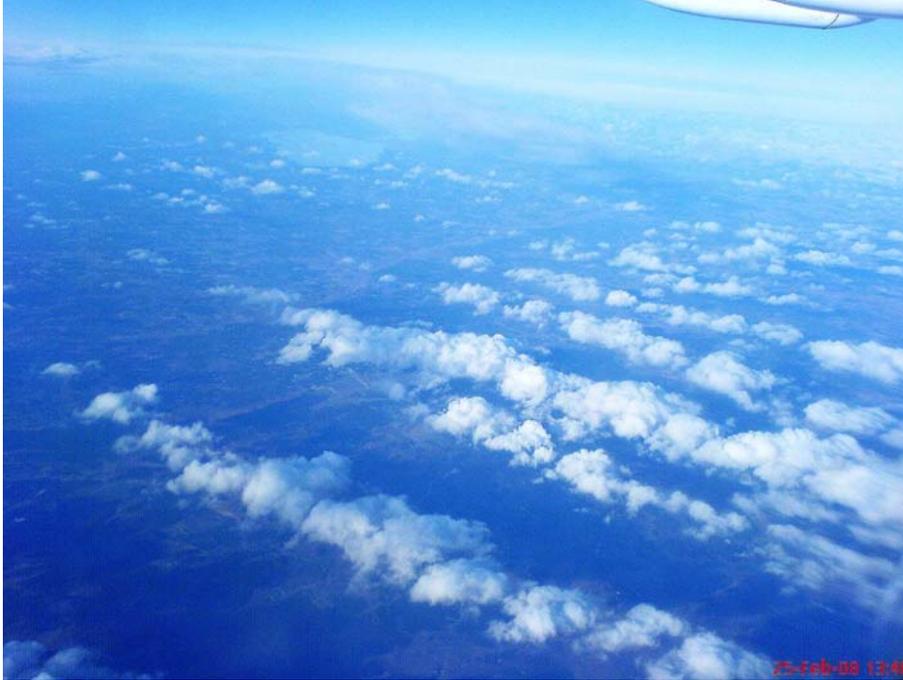

**Fig. 6:** Large-scale turbulent structure pattern known as rolls or cloud streets over southern Finland. Photo taken from altitude of 8 km height 25 February 2008 by I. Esau.

Weinbrecht et al. (2004) compared LST near the surface in the convective LES and those derived from acoustic tomography. This method provides two-dimensional data arrays, which are probably more suitable for LES-validation because the tomographic data are area- or volume-averaged. This study highlighted practically important, but to some degree unexpected, sensitivity of the LST to small details and variability of the external parameters such as surface heat flux variations. This showed that for determining some boundary conditions, e.g. the surface temperature and the roughness length, high measurement accuracies are necessary, which are difficult to reach or which require considerable extra measurement efforts. The unpleasant conclusion of this study was that the standard accuracy of the meteorological measurements is insufficient for reasonable set up of LES and intercomparisons with observations. However, a good qualitative agreement with some quantitative differences has been demonstrated.

Scipion et al. (2008) compared observed radar radial velocities with LES. The minimum averaging time required for the signal to approximate spatial ergodicity was estimated to be approximately between 30 s and 120 s for the idealized CBL case. Longer averaging produced decorrelation due to the breakdown of statistical stationarity in the signal. Good qualitative agreement was found in the zonal and meridional wind fields retrieved from the radar and estimated with LES. A somewhat larger disagreement was found for vertical velocities. This disagreement may be attributed to an instrument error. After removing the bias, the PDFs from the radar and realistic LES cases were in agreement.

Summarizing, it has been demonstrated qualitative utility of LES in geophysical applications but the quantitative accuracy of LES has not been verified yet. To my opinion, the major obstacle on the way is absence of a kind of data assimilation routine working on turbulence scales. Such a routine could improve the LES comparability a lot (Cheng et al., 2001). Certain steps in this direction are undertaken (Carrio et al., 2006) but the experience is yet minor. So far, LES based conclusions, if not speculative, are mostly based on convergence and similarity arguments. Namely, it is argued that LST features reproduced in a large variety of runs with different resolution and slightly modified external conditions are robust and should correspond to the realistic LST features.

## 3. EKMAN BOUNDARY LAYER

The simplest example of a realistic geophysical PBL is the Ekman boundary layer (EBL). The EBL develops in a boundary layer flow, not necessary turbulent flow, in the rotated frame of reference. Being the closest observed analogy to the laboratory flows, the EBL has been intensively studied over the last 50 years. Still, LES have introduced a new value of these studies. The equilibrium EBL over aerodynamically rough but flat homogeneous surface is controlled by the single external non-dimensional parameter, the surface Rossby number $\mathrm{Ro} = U_g / f z_0$ or its variants (Esau and Zilitinkevich, 2006). Here, $U_g$ is the geostrophic wind speed above the EBL, $f$ is the Coriolis parameter and $z_0$ is the roughness length scale for momentum. For concrete definitions of those parameters see e.g. Garratt (1992). We will look here at three principal moments of geophysical LES applications, which cause major difficulties. The first one is the surface boundary conditions and the surface layer. The second one is an effective Re. The third one is LST self-organization.

**The surface layer**

The surface layer is a layer where the vertical profile of the mean velocity, $\overline{u}(z)$, is approximately logarithmic function of height. The surface layer has been acknowledged as a significant problem for the LES applications (Mason, 1994) due to the simple reason that the turbulence scale is proportional to the height in this region of the flow. In fact, the log-law of the wall

$$\overline{|\tau|^{1/2}} = \overline{u_*} = \frac{\kappa \overline{|u(z)|}}{\ln(z/z_0)} \qquad (7)$$

was obtained with Eq. (5) as the assumption[2]. Here $|\tau| = u_*^2$ is the momentum flux and the surface friction velocity, correspondingly. It is also assumed that the flux is height independent within the surface layer. The pointwise and instant application of Eq. (7) is usually used as the surface boundary conditions in the geophysical LES[3].

---

[2] Other methods of derivation of the log-law of the wall are possible (see Oberlack, 2001).

As the mixing length $l_s(z)$ decrease with $z \to 0$, the turbulence becomes severely under-resolved at the first few computational levels in LES. To some degree, this problem can be alleviated with the variational formulation of the turbulence closure (Porte-Agel et al., 2000) but arguably it cannot be avoided entirely. At the same time, it is known that the surface layer is the layer where the turbulence is generated. This knowledge came, and it is important to emphasize, from the single-point analysis of the mean TKE budget from observations (e.g. Caughey and Wyngaard, 1979). On this basis, several researchers questioned reliability of the geophysical LES as LES cannot resolve the turbulence generation process. Indeed, earlier LES were too coarse and too over-dissipative to perform reasonably. In the intercomparisons by Andren et al. (1994), LES codes were unable to simulate the non-dimensional velocity gradient, which must be $\phi_m = 1$ in the EBL, with accuracy better than 20%. The LES has improved very much over last years. Figure 7 shows $\phi_m$ in two recent EBL simulations with LESNIC. As one can see the anomalies in $\phi_m$ are less than 5% on average. The runs allow determination of the optimal values of the von Karman constant ($\kappa = 0.39$ to $0.46$) as well as the thickness of the log-layer ($z_s = 0.25\,h$).

Although opponents questioned LES utility, proponents pointed out on surprisingly robust performance of the EBL LES in numerous experiments. Satisfactory but still somewhat qualitative explanation of the LES performance in the surface layer appeared from observations of the air-sea interactions. Figure 8 illustrates that under sufficiently strong wind the flow-to-surface interaction can be visualized through imaging of short-lived (life time 1-5 s) capillary waves. On larger scales this effect is used to explore sea surface stress variability with the satellite aperture radar systems. As it is seen, the maximum stress patches or the spots with rough and therefore dark surface are not organized in a hierarchy of scales as the Prandtl view would suggest. Contrary, there are chaotic but of rather certain scale imprints on the surface. Fortunately for LES, the horizontal scale of the imprints is rather large $O$(100 m). It suggests that a significant if not major part of the flow-surface interaction is carried on by the descending LST. This is true at least for high Re geophysical PBL. Hunt at al. (Hunt and Morrisson, 2000) developed a theory of such LST-based interactions that has been partially corroborated through observation analysis by Hoegstroem et al. (2002).

Summarizing, we can conclude that geophysical LES is a robust and instructive tool to study the surface layer provided the required correction of the energy cascade are taken into account through the turbulence closure. Again, this correction is not necessary to be taken through the dynamic eddy-viscosity model albeit other approaches are yet to demonstrate their efficiency and universality. The key for its robustness is fortunate large-scale organization of the flow-surface interactions in high Re PBL where the control agents, the LST, can be easily resolved in LES on feasible grids.

---

[3] In stratified PBL simulations, LES utilize the surface boundary conditions based on Monin-Obukhov functions in addition to the log-law of the wall. Their discussion is out of scope of this overview.

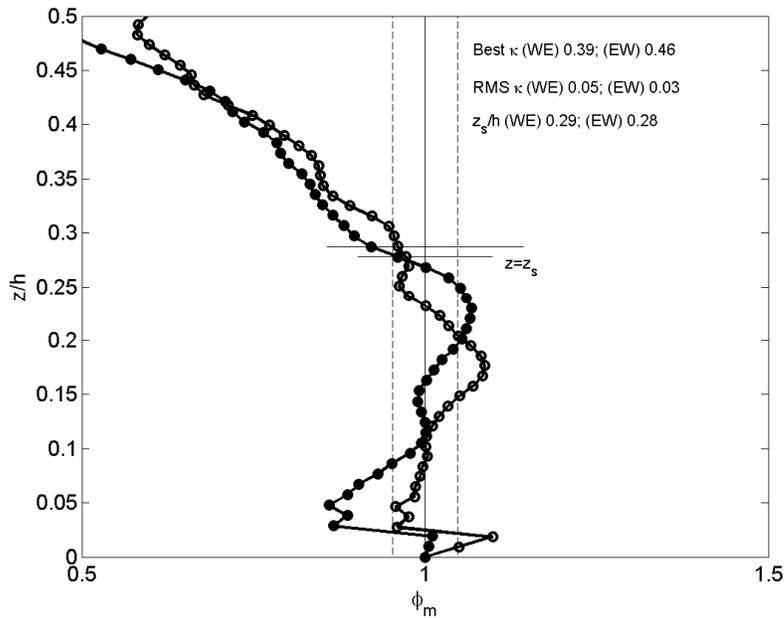

**Fig. 7:** Non-dimensional velocity gradient $\phi_m$ as function of the height $z$ normalized on the EBL depth $h$. The data are results from two LESNIC runs for the EBL at the latitude 45N with west-east (WE) and east-west (EW) mean winds of 10 m s$^{-1}$. The run resolution was 256 by 256 by 128 nodes in the domain 9 km by 9 km by 6 km in zonal, meridional and vertical directions correspondingly.

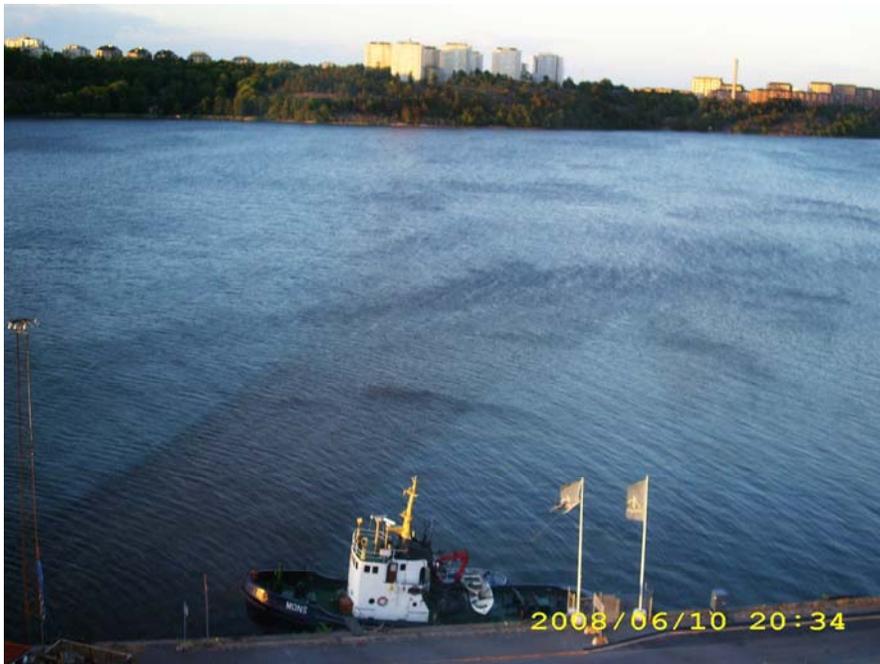

**Fig. 8:** Observation of the flow-to-surface interaction under sufficiently strong wind of 15 m s$^{-1}$. The interactions are visualized through imaging of short-lived (life time 1-5 s) capillary waves seen here as darker rough spots with relatively large surface stress. Photo taken from Scandic Ariadne Hotel in Stockholm by I. Esau.

## The effective Reynolds number

One of the deepest philosophical problems related to geophysical LES is the correspondence between numerical or LES-fluid and the geophysical fluids. The molecular viscosity of air and water is small, $10^{-5}$ m$^2$ s$^{-1}$, that for any practical reason it can be neglected. Instead, as Muschinski (1996) posed it, one has to deal with LES-fluid that is fluid with non-linear viscosity that has mathematical properties of the advection operator $u_i \partial/\partial x_j$ from which it has been originated. There are two fundamental questions rising. Does this LES-fluid have the same structural properties in the sense of statistical fluid mechanics as geophysical fluids? If so, what is the effective Re of the LES-fluid flows?

Indeed, different useful universal scalings on Re and corresponding wall units are established in fluid mechanics of the bounded turbulent flows such as the PBL. The most useful with respect to geophysical flows is the universal Re-independence of the turbulence at sufficiently high Re. Wei and Wilmarth (1989) noted such a behavior in laboratory flows at Re ~ $O(10^4)$ and Coleman et al. (1990) in direct numerical simulations of the EBL and their theoretical approximation. If the Re-independence is reached, there is no need for higher Re experiments. For the LES applications where Re and the amount of grid nodes are proportional (see Introduction), Re-independence translates into the results convergence toward Re $\to \infty$ statistics and straightforward into computational resources.

Fortunately, we can infer some estimates of the effective Re in LES using sensitive Re-dependent relationships. It is worth noticing that many published estimations of the effective Re (e.g. Cai et al., 1995) are irrelevant to the problem. Such estimates are based only on the number of grid nodes but do not account what the model computes on that mesh. More relevant but also phenomenological estimate has been proposed in Esau (2004). Here, we consider an advanced theoretical construction to infer the effective Re in LES.

To move on, one should observe that the Oberlack (2001) symmetry analysis provides two possible scaling laws for the surface layer in the PBL. Since the surface layer is the worst resolved layer in LES, it should have the lowest restrictive Re in simulated flow. Except the log-law of the wall, this layer possesses also an exponential law of the wall proposed by Barenblatt et al. (2002). It takes the form

$$U/u_* = A\left(\frac{u_* \operatorname{Re}}{U} \frac{z}{h}\right)^a, \qquad (8)$$

where $A = 3^{-1/2} \ln \operatorname{Re} + 5/2$, $a = 3/2 \ln \operatorname{Re}$ and an effective Re is a free constant to fit the data. Figure 9 demonstrates this fit with EBL data obtained with the LESNIC code. The best fit was found for the effective Re ~ 64000 which is well in the Re-independent range but still much less that usually assigned Re for geophysical LES.

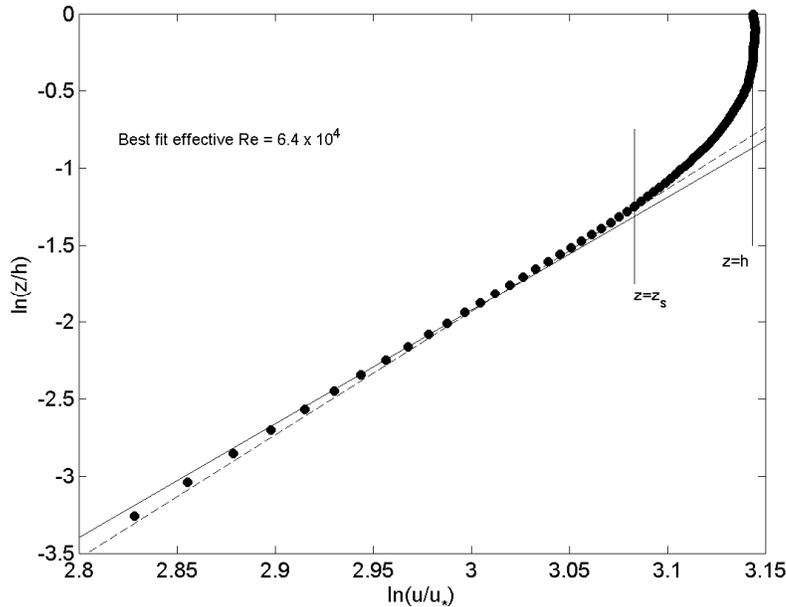

**Fig. 9:** The surface layer scaling in log-log coordinates: dots – LESNIC EBL data for WE run (see Fig. 7); solid line – the exponential law after Eq. (8); dashed line – the traditional log-law after Eq. (7).

**Self-organization of turbulence in EBL**

Simplicity of the EBL allows deeper look at the turbulence self-organization and its role in the total turbulent transport. The mean Ekman wind profile is unstable with respect to small perturbations, as it is known since Lilly (1966). But the linear stability analysis of the high Re three-dimensional EBL is hard to prove. Self-organized turbulence in the PBL is usually observed under convective conditions (see Fig. 6). Evidences for the EBL LST were less certain. Mason and Sykes (1980) used two-dimensional LES to resolve the question. They identified predicted inflection point instability in their LES. However, the model cross-flow domain size (10 km) accommodated only one or two counter-rotating roll pairs. It raised doubts whether these structures an artifact of the periodic boundary conditions they used. Later, Mason and Thompson (1987) failed to identify rolls in their three-dimensional LES. Coleman et al. (1990) also found no rolls in their DNS. Those later LES and DNS experiments were conducted in too small domains as well.

Taken proper cross-flow domain size (0.5 km by 144 km by 6 km) the structures were simulated in LESNIC experiments. Figure 10 shows sections (~ 1/10 of the domain size) in two experiments with and without LST. Figure provides evidences if not prove that the LST does self-organize in adequate LES and likely does so in the real EBL. The role of LST is large. In this LES, LST practically double the momentum flux and increase the TKE by the factor of 5 in the EBL. In more realistic runs, the effect is expected to be significantly less but still of major importance. This large effect is due to the large scale of the emerging LST. As one can see the horizontal scale of an individual roll is about 2-3 km and the inter-roll scale is about 4-5 km. Somewhat larger scales are also observed emerging in LES.

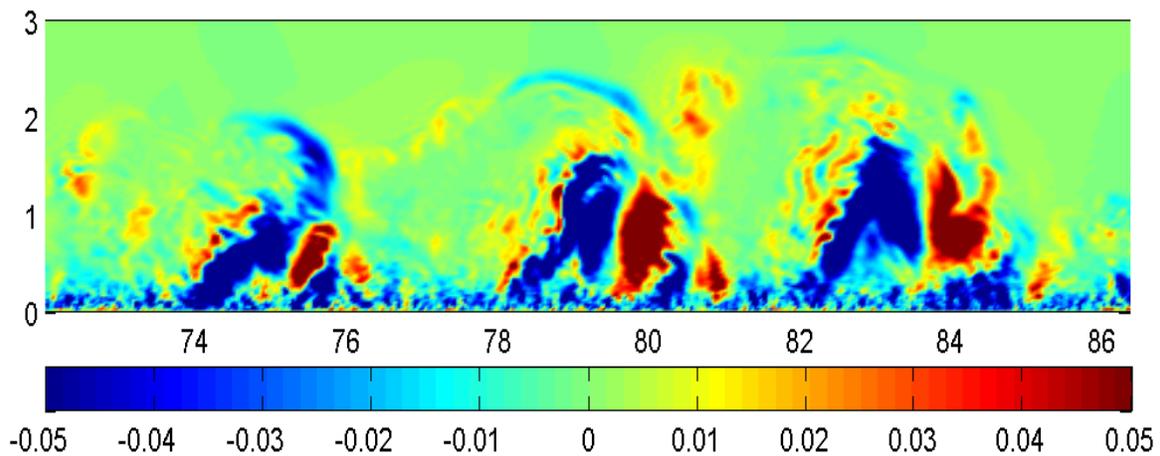

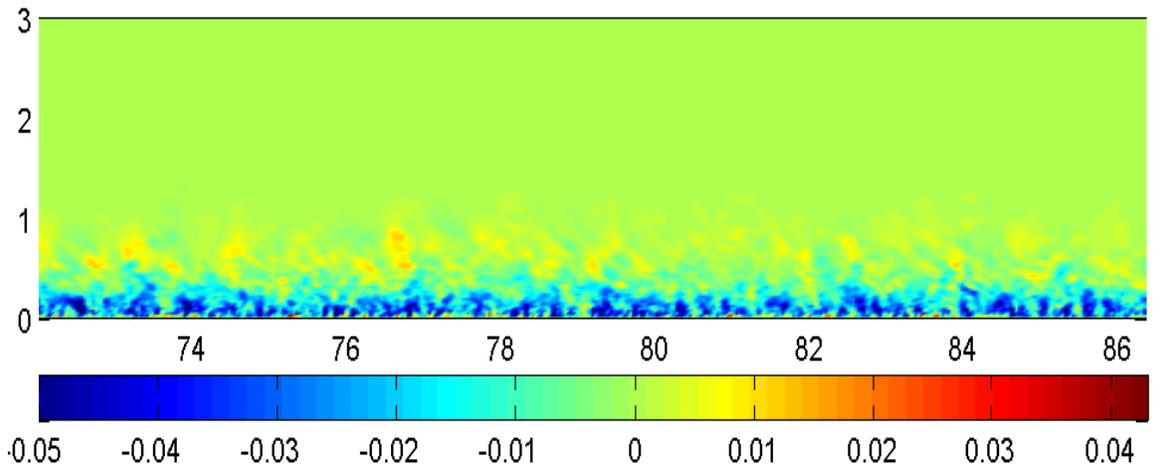

**Fig. 10:** The large-scale self-organized turbulence in form of streamwise roll structures visualized through the vertical momentum flux (red – upward, blue –downward) in two LESNIC runs. The roll axis is perpendicular to the cross-section shown. The model was run under the same conditions controlling the EBL except orientation of the mean flow with respect to the LESNIC domain: (a) the geostrophic wind to latitude (west-east) angle is 18° north as it has been predicted for the most unstable mode (e.g. Brown, 1972); (b) – the same but for the orientation of 45° south, predicted for the least unstable mode. The color scale is for the vertical momentum flux in m$^2$ s$^{-2}$. The vertical and horizontal scales are in km. The cross section is oriented along meridian and centered at 45N latitude.

## 4. APPLICATIONS OF GEOPHYSICAL LES

Unfortunately even an overview presentation does not survey all essential applications of the geophysical LES. Figure 11 shows the growth of the LES field measured as numbers of publications with the term "large eddy simulation" in titles. Obviously, this is only a fraction of all publications, especially applied ones, where LES results have been used. My inside into literature in some specific directions of interest, e.g. polar PBL studies and LST structure studies in geophysics as well as the von Karman constant and the Smagorinsky coefficient studies in fluid dynamics, suggests that geophysical applications use LES data about 5 to 8 times more often than they use the term in their title. This is not surprising as applications are problem-oriented unlike fluid dynamics studies, which are method-oriented. For fluid dynamics studies I would suggest the multiplication factor 3 to 5. So in 2008, LES has been used in 450-600 published works totally and likely in 100-150 works in the area of geophysics.

In view of the abovementioned diversity, this overview does not present a collection of summaries from those relevant publications. This overview highlight only major, to my subjective opinion, problems resolved or to be resolved in geophysical LES.

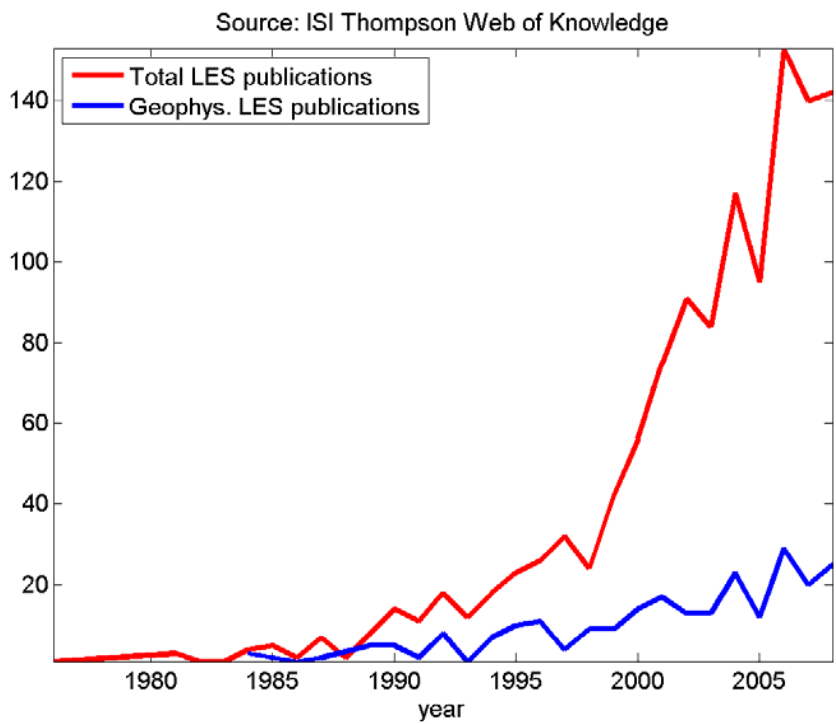

**Fig. 11:** Number of publications appeared each year with the term "large eddy simulation" in titles. Total amount of publications (red curved) includes all scientific fields and the amount of geophysical publications (blue curve) includes only meteorology, oceanography and air quality publications. Large increase of publications between 1997 and 2000 likely indicates expansion of the LES research on increasing desktop PC computer power. Source: ISI Thompson Web of Knowledge.

**LES of convective PBL**

It has been mentioned in Section 2 that LES of the convective PBL were generally rather successful and helpful. This success is to be attributed certainly to the large scales of the energy-containing sub-range and the wide inertial sub-range with its upper limit often at a few hundreds meters. The latter feature makes it easy to resolve the major share of the turbulence energy even in coarse resolution LES. Moreover, the long inertial sub-range ensures a large degree of the turbulence isotropy since the additional energy input due to the work of buoyancy forces is located at the large scales.

Although the simulations of the convective core of the PBL are of no problem for LES even at coarse resolutions (e.g. Deardorff, 1980; Khanna and Brasseur, 1997), the surface layer and the layer of capping temperature inversion at the PBL top are. Otte and Wyngaard (2001) concluded on the bases of 26 high-resolution large eddy simulation runs ranging from neutral, inversion-capped to free-convection cases that the stable stratification induced by the capping inversion is of dominant influence on entrainment and counter-gradient fluxes into the PBL.

Another problem, generally overlooked in earlier studies, is the large scale and equally large importance of the LST self-organization in the convective PBL (Schumann et al., 1989). Figure 6 shows the LST organized in rolls under strong wind. Figure 12 shows the LES organized in cloud cells under slow wind. According to Barthlott et al. (2007), effect of coherent structures is recognizable during 36% of the total time of long-term atmospheric observations in the surface layer. The LST contribution is on average of 44% in the momentum flux and 48% of the heat flux. So the LST transport heat more efficiently than momentum making the turbulent Prandtl number less than one, which is in good agreement with observations (see below). The LST have larger significance over smooth surface (Zilitinkevich et al., 2006) where they are the primary drivers for the vertical turbulent exchange.

Obviously, since the LST play such a big role in the convective PBL, at least several LST structures have to be resolved in the LES domain. It imposes severe lower limit on the domain size. Indeed, Rayleigh-Benard convection has been thoroughly studied in laboratories. Its minimum aspect ratio $h:\lambda$ at large Rayleigh numbers is well known to be about 1:2. Thus the horizontal size of the LES domain has to be larger than about $6h$ to accommodate at least 3 LST structures. This minimal requirement demands the domain of 9 km to 12 km and possibly larger. In fact, the geophysical LST aspect ratio is larger and often reaches 1:4 and more (Müller and Chlond, 1996; Atkinson and Zhang, 1996; Müller et al., 1999). In this case, the horizontal domain size increases to 15 km to 25 km and more. These conditions have been practically never respected in LES runs. Thus conclusions of the absolute majority of publications based on those LES runs must be treated with caution. Done properly however, the bulk properties of the convective PBL compare reasonably well with observations[4]. Figure 12 presents the LST pattern in a LESNIC run for a convective shear-free PBL.

It is now well understood that details of the convective PBL are equally computationally demanding as the details of the shear-driven PBL. It is needed to resolve in LES both the largest scale of the LST and the relatively small-scale of convergence or squall lines between them. The ratio of those scales can reach 100, which imply the demand of 512 by 512 node mesh in horizontal direction. Such LES runs are feasible on super-computers with PALM code[5] (IMUK, University of Hannover, Raasch and Schroeter, 2001) and NCAR code (NCAR, UCAR, USA, Sullivan et al., 2008). Sullivan and Patton (2008) reported convective PBL LES runs up to $1024^3$ grid nodes in 5 km domain. This LES used 8192 processors of a Cray XT4. Flow visualization and statistics are compared for resolutions varying from $100^3$ to $1024^3$. For the simulated case, an excellent convergence of results has been reached already at $256^3$ resolution. At the same time, the LST structures remained unresolved in the chosen domain.

---

[4] It needs to be reminded that the accuracy of atmospheric PBL observations are less than those in laboratory and that the external parameters are usually known only approximately.

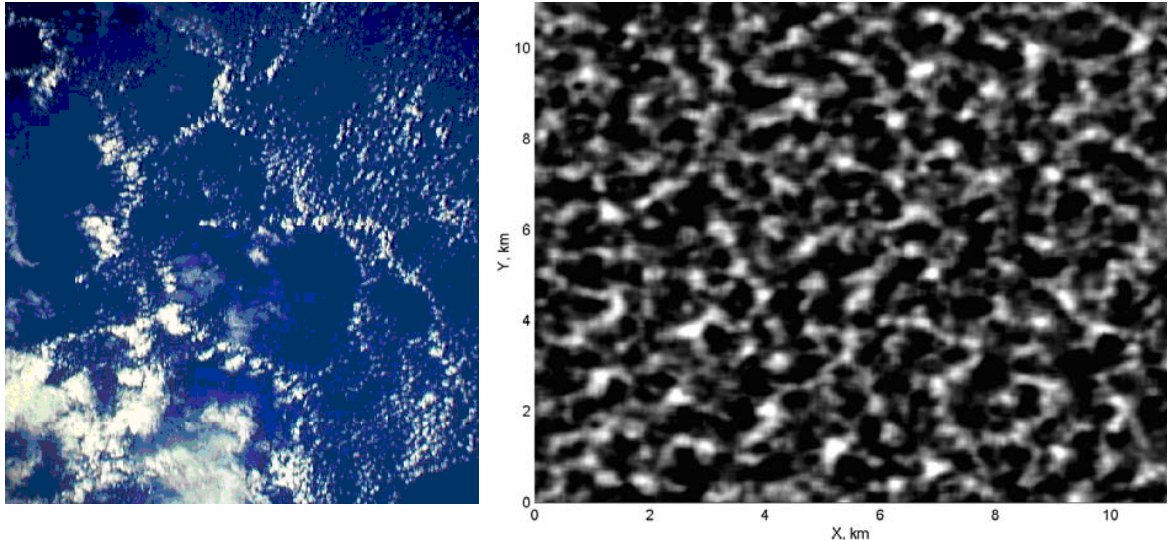

(a)          (b)

**Fig. 12:** (a) Cloud cells over Pacific Ocean (26.06.1992; 22:41:17 GMT Time; Nadir point latitude: 11.6, longitude: -118.2). Mission: STS050; Roll: 72; Frame: 65 Mission ID on the Film or image: STS50. Image courtesy of the Image Science & Analysis Laboratory, NASA Johnson Space Center. Reprinted in compliance with "Conditions of Use of Astronaut Photographs" (http://eol.jsc.nasa.gov). (b) Convective cells in LES as visualized with turbulent kinetic energy (dark areas correspond to lower energy content) at $0.2\,h$. Simulations with LESNIC code were computed for Zilitinkevich et al. (2006).

**LES of stably stratified PBL**

Contrary to convective PBL LES, LES of the stably stratified PBL (SBL) became possible only in the last decade. In 2000, Saiki et al. (2000) reported failure of LES of weakly stratified SBL if it runs without special precautions. The first successful LES should be probably attributed to Kosovic and Curry (2000) who used a new non-linear turbulence closure (Kosovic, 1997).

The central problem in the SBL studies is related to the mechanical energy conservation (Zilitinkevich et al., 2008). Following the Kolmogorov's analysis done to the EBL, only one component of the mechanical energy of turbulence, namely, the turbulent kinetic energy (TKE) has been considered. It led to the erroneous conclusion that TKE is irreversibly dissipated either directly through the energy cascade or indirectly through the work against buoyancy forces in the SBL. It has been overlooked that the work against buoyancy forces is reversible. In fact, the TKE is not dissipated in this process but stored as the turbulent potential energy (TPE). Part of the TPE can be converted back into the TKE through the internal gravity wave radiation. The equation for the total, i.e. conserved in the absence of dissipation, turbulent energy (TTE), $E$, is

$$\frac{DE}{Dt} + \frac{\partial \Phi_E}{\partial z} = -\boldsymbol{\tau} \cdot \overline{\mathbf{S}} - \varepsilon_E. \qquad (9)$$

---

[5] The code PALM is now installed at the Nansen Environmental and Remote Sensing Center, Bergen, Norway

Here, $\varepsilon_E = \varepsilon_K + \varepsilon_P$ and $\Phi_E = \Phi_K + \Phi_P$ are the dissipation rate and the vertical turbulent flux of the TTE as the sums of corresponding quantities for the TKE and the TPE respectively.

Maintaining turbulence at large Richardson number, Ri, is achieved as follows. Suppose that the buoyancy flux becomes so large that TKE considerably decreases. The TTE is conservative quantity, therefore the TPE increases and fluctuations of buoyancy strengthen. In other words, fluid elements acquire stronger accelerations and speed up toward their "equilibrium level", which causes re-establishing of TKE, and decreasing of TPE. In its turn, too large TKE causes stronger displacements of fluid elements, hence stronger buoyancy fluctuations and therefore increasing of TPE. Such oscillations are typical of intermittent turbulence where waves are radiated in one place and break up with mixing elsewhere.

The proposed wave radiation mechanism is plausible but yet to be proved. It however is in good agreement with available data. In particular, it explains the behavior of the mixing intensity or the Prandtl number, Pr, at large Ri. Figure 13 demonstrates this behavior. The consequences of the new TTE theory are not completely understood yet. Its development has been based largely on the SBL data where the non-local effects due to the TKE vertical transport from the surface upward are significant. In fact, large Ri are observed only at the SBL top where turbulence is already weak and intermittent.

**LES over heterogeneous**

So far only LES applications over homogeneous surface have been considered as the progress in that direction has advanced the most. Those simulations are more relevant to water areas far offshore. Their importance is determined by the fact that the Earth's surface is the ocean by 70% and ice sheets and lakes by another 5%. The remaining 25% of the realistic surfaces over land are always heterogeneous.

LES over flat but thermally or roughness heterogeneous surface are almost equally complicated as those over homogeneous surface. The analysis of results is complicated. The surface heterogeneity imposes an external scale, $\lambda$, of the fluid forcing. This scale can be found in different parts of the turbulent spectra. Moreover, it may have complex spectral structure itself.

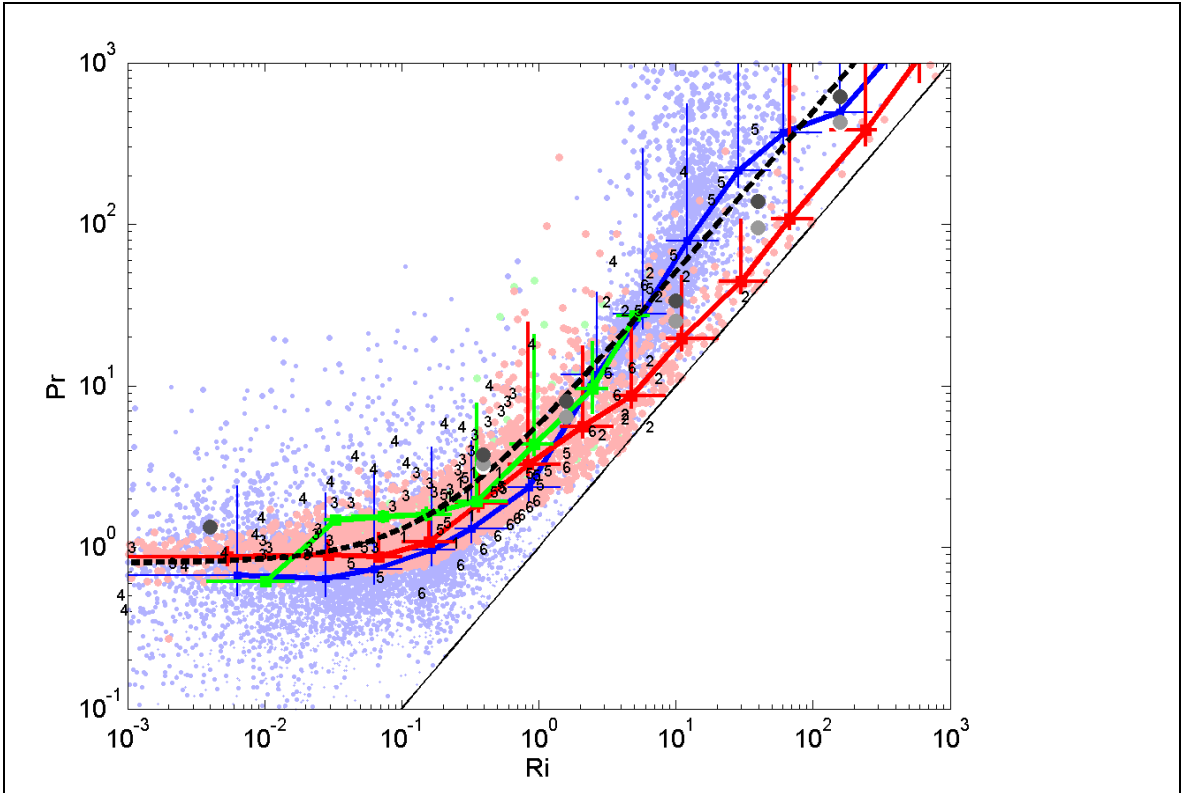

**Fig. 13:** Turbulent Prandtl number $Pr_T = K_M / K_H$ versus Ri. Blue points and curve – meteorological campaigns SHEBA (Uttal *et al.*, 2002, mostly for Ri < 1) and CASES-99 (Poulos *et al.*, 2002, for 0.1 < Ri <100); green – laboratory sheared flow (Ohya, 2001); red – new LES using LESNIC code (Esau 2004); grey – direct numerical simulations (DNS) with 32 (lightest), 64 (darker) and 128 (darkest) nodes, respectively (Stretch *et al.*, 2001). Numbers show data from literature (for details see Zilitinkevich et al., 2008). The dashed curve: Pr = 0.8 + 5Ri is composed of the two asymptotes: already known: Pr = 0.8 at Ri < 0.1, and obtained from this figure: Pr = 5Ri at Ri >1. Red, green and blue curves show bin-averaged data for the corresponding data sources. Horizontal bars show the width of bins. Vertical bars show one standard deviation above and below the averaged value within the bin. The thin line: Ri/Pr = $Ri_f$ = 1 separate the "principally impossible area" ($Ri_f$ cannot exceed unity in the steady state).

At present the problem is studied in terms of idealized linear (Esau, 2007) or check-board arrangement of heterogeneities. Some degree of understanding has been gained from analysis of $\lambda / h$ ratio under different conditions (Patton et al., 2005; Esau, 2007; and references therein). In particular, it has been recognized that for $\lambda < h_b << h$, the role of surface heterogeneity is minor. The vertical scale $h_b$ or the blending height (Claussen, 1995) was introduced to separate the layer of local turbulence equilibration from the PBL as such. If $h_b < \lambda < h$, the heterogeneities have the major effect on the amplitude and phases of the LST causing significant modification of the turbulent fluxes, statistics and the LST themselves. Nevertheless within this range of scales, the PBL still has to be considered as a forced system with non-linear response. This is the most interesting, very frequently observed and at the same time the least understood range of scales. If $\lambda > h \gg h_b$, the PBL breaks up on more or less independent set of individual equilibrium PBLs with locally homogeneous properties. Figure 14 illustrates the surface heat flux magnitude in all three $\lambda$ ranges.

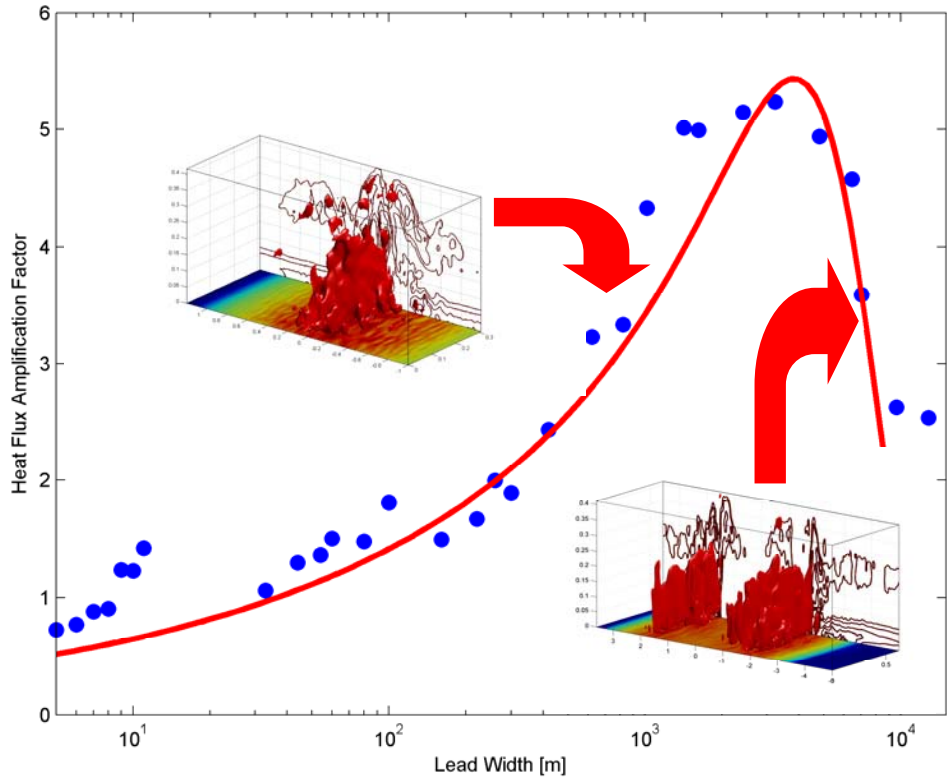

**Fig. 14:** Magnitude of the surface sensible heat flux amplification factor, $H_s(\lambda)/H_s(\lambda \to \infty)$, as function of the surface heterogeneity scale $\lambda$ in LESNIC experiments with temperature variations due to opening of Arctic sea ice leads (Esau, 2007). Red curve is an empirical fit. Blue dots are LES runs with fixed $\lambda$ and the surface temperature difference of 8 K. Insiders visualize the heat flux (color surface) and the turbulent kinetic energy (black contours) of the flow corresponding to $h_b < \lambda < h$ (upper left) and $\lambda > h \gg h_b$ (lower right) scale ranges.

## 5. SUMMARY AND DISCUSSION

In the summary and discussion Section, the following moments should be brought to the readers' attention. First, robustness of the geophysical LES has improved enormously after introduction of the to dynamic turbulence closures. The reason for this improvement is more accurate local and instant representations of the spectral energy cascade in the inertial sub-range of scales. In spite of this, the absolute majority of the LES codes (Beare et al., 2006) still employ non-dynamic closures with some adjustments in a few cases

(e.g. Sullivan et al., 1994; Kosovic, 1997). It leads to the need for finer, more computationally demanding mesh in the geophysical LES.

Second, it has been understood, albeit still incompletely, that the high Re PBL interacts with the surface in a somewhat different way than its low Re counterpart. It is recognized that the low Re PBL interacts with the surface through sweep-ejection events (e.g. Robinson, 1991). The high Re PBL is different in a sense that the sweep-ejection events are related to descending large eddies from the PBL core. This observation makes geophysical LES surprisingly robust even in the under-resolved surface layer.

Third, feasibility of LES with mesh resolution of $O(10$ m$)$ or less allowed studies the local turbulent processes in the PBL in fine details. Their qualitative and quantitative theory has been now generally constructed including convective and stably stratified cases. An important moment here is that there is no need to study geophysical turbulence at arbitrary high Re. The problem was only to reach sufficiently high Re where the turbulence properties become Re-independent. This could be achieved at Re ~ $O(10^5)$. Similar notions can be made about turbulence behavior with respect to Ri.

Fourth, it was understood that the geophysical PBL turbulence is always self-organized. The scale of its self-organization may be in different relation to the externally imposed scale of spatial and time variability. The scale ratio causes strong effects on the PBL turbulence.

Fifth, many important applications of the geophysical LES remained outside this overview. To mention a few without ranking their significance, there are LES of stratocumulus clouds and cloud-turbulence interactions (e.g. Stevens et al. 2005; Chlond et al., 2004). Rather advanced studies in this direction were conducted and excellent results achieved by Peter Dyunkerke (Holtslag and Duynkerke, 1998). There are LES of the ocean mixed layer (e.g. McWilliams et al., 1997; Skyllingstad and Denbo, 2001; Noh et al., 2004). At larger scale, there are simplified LES used as a kind of super-parameterization in the cloud-resolving models including global circulation and climate model applications (Wyant et al., 2006) as well as so called very large eddy simulations.

Sixth, several studies conducted intercomparisons between geophysical LES and observed turbulence in the PBL. The studies noticed rather well qualitative and up to some degree of quantitative statistical agreement between the LES and the data. At the same time, it was difficult to establish complete quantitative correspondence primary because of imperfections of the observation techniques but also because of inability to run LES under the observed PBL conditions.

Finally, we shell highlight the front edge of the geophysical LES research. Although the local, column-wise turbulence properties are rather well studied, the properties of the non-local turbulence and turbulence self-organization on the PBL and larger scales are barely touched. There are very few works with properly done simulations and quantitatively supported conclusions (e.g. Gryshka et al., 2008; Sullivan and Patton, 2008). It is not surprising as only in the most recent years massive-parallel computers made it possible to run adequate LES. Simulations with mesh resolutions of 2048 x 2048 x 1024 nodes have been already presented by both IMUK and NCAR groups. It is not only the question of the computer time availability but also storage and analysis of such a large data massive cause problems.

Among other directions of research to be pursued, data assimilation into geophysical LES and formulation of the surface boundary conditions with under-resolved surface geometry are to be mentioned. Arguably, any move toward realistic, not to say predictive, simulations requires LES runs under observed conditions, which are known inaccurately and incompletely. In geophysics the problem is solved through assimilation of the available observations into the large-scale models. The assimilation forces the model to minimize the difference between the model and observations in given metrics on the given sub-set of grid nodes. Similar procedure has to be invented into LES. Prospective research in this direction has been published by e.g. von Storch (2000). The boundary condition problem is in some sense similar to the assimilation problem. The difference is that the conditions are known, e.g. surface topography is known everywhere, but cannot be approximated with a coarse LES mesh. Prospective research in this direction has been published by Wunsch (2003).

Concluding this overview, I want to say that geophysical LES is rapidly evolving field of research, which is already just a step away from practical applications. If earlier LES has to be carried on a few super-computers, many of the recent studies can be done on consumer level PC. There is no doubt that over the next few years we will find a broad spectrum of interesting and helpful LES studies published.

**ACKNOWLEDGEMENTS**

This work has been supported by the EU FP7 projects ERC PBL-PMES (No. 227915) and MEGAPOLI (No. 212520), the FRINAT-Norway project PBL-FEEDBACK, and the Norwegian Research Council projects PAACSIZ, POCAHONTAS and NORCLIM.